\documentclass[journal,twoside,web]{ieeecolor}
\usepackage{lcsys}
\usepackage{graphicx}
\usepackage{cite}
\usepackage{amsmath,amssymb,amsfonts}
\usepackage{textcomp}
\usepackage[hidelinks]{hyperref}

\def\BibTeX{{\rm B\kern-.05em{\sc i\kern-.025em b}\kern-.08em
    T\kern-.1667em\lower.7ex\hbox{E}\kern-.125emX}}
\newtheorem{assumption}{Assumption}

\newtheorem{lemma}{Lemma}
\newtheorem{theorem}{Theorem}
\newtheorem{remark}{Remark}

\newtheorem{problem}{Problem}

\DeclareMathAlphabet{\mathbbb}{U}{bbold}{m}{n}
\newcommand{\col}{\operatorname{col}}
\newcommand{\diag}{\operatorname{diag}}

\newcommand{\argmax}{\operatorname{argmax}}

\newcommand{\sat}{\operatorname{sat}}

\begin{document}
\title{Data-driven Dynamic Intervention Design in Network Games}
\author{Xiupeng Chen, and Nima Monshizadeh, \IEEEmembership{Member, IEEE}
\thanks{Xiupeng Chen and Nima Monshizadeh are with the Engineering and Technology Institute, University of Groningen, 9747 AG
Groningen, The Netherlands.  (e-mail: \{xiupeng.chen, n.monshizadeh\}@rug.nl).}
}

\maketitle
\thispagestyle{empty}

\begin{abstract}
Targeted interventions in games present a challenging problem due to the asymmetric information available to the regulator and the agents. This note addresses the problem of steering the actions of self-interested agents in quadratic network games towards a target action profile. A common starting point in the literature assumes prior knowledge of utility functions and/or network parameters. The goal of the results presented here is to remove this assumption and address scenarios where such a priori knowledge is unavailable.
To this end, we design a data-driven dynamic intervention mechanism that relies solely on historical observations of agent actions and interventions. Additionally, we modify this mechanism to limit the amount of interventions, thereby considering budget constraints. Analytical convergence guarantees are provided for both mechanisms, and a numerical case study further demonstrates their effectiveness.
\end{abstract}


\section{Introduction}
\IEEEPARstart{N}{etwork} games have become a powerful tool for analyzing the strategic interactions of agents in various societal and economic systems, including crime networks \cite{ballester2006s}, social networks \cite{bloch2013pricing}, and public goods provision \cite{bramoulle2007public}. A key assumption in network games is that each agent's payoff depends not only on their own actions but also on the aggregated actions of the agents in their neighborhood \cite{parise2019variational}. These interactions among agents are modeled by an underlying network, represented by a graph.

Since agents in network games are self-interested and solely focus on maximizing their own payoffs, the outcomes of such strategic interactions are often suboptimal, leading to a degradation in the overall system's performance. This inefficiency is typically quantified by the price of anarchy \cite{koutsoupias2009worst}. Therefore, with a deep understanding of these strategic interactions within the network game framework, there has been significant interest in influencing outcomes through efficient interventions, aiming to align individual actions with more desirable global outcomes. For instance, in \cite{galeotti2020targeting, parise2021analysis}, a central regulator intervenes by modifying the standalone marginal benefits of agents to achieve social welfare optimization. It is demonstrated that the optimal intervention policy is highly dependent on the parameters of utility functions and the underlying interaction network. However, in practice, due to the scale of the system and privacy concerns, the underlying interaction networks often remain hidden for the regulator, and the utility functions of agents are not readily available. This lack of information poses significant challenges for designing effective interventions.  

Control-theoretic tools have been employed to enhance the performance of systems exhibiting self-interested behavior under information constraints \cite{alpcan2009control}. In this setting, agents iteratively adjust their actions through game learning dynamics in response to the regulator's interventions. Simultaneously, the regulator observes these actions over time and updates its interventions accordingly. This situation can be conceptualized as a feedback control problem, where the central regulator formulates suitable control laws to steer the agents' actions toward a desired outcome. In \cite{eksin2020control, riehl2018incentive}, two incentive-based control methods were proposed to steer agents in network games toward the desired equilibria. However, these methods are not applicable to network games with continuous action spaces. For more general games, a model-based algorithmic framework was developed in \cite{ratliff2020adaptive} for adaptively computing interventions without requiring knowledge of game learning dynamics. In \cite{maheshwari2024adaptive}, dynamic interventions are updated based on both the agents' marginal costs and the society's marginal costs. The analyses in \cite{ratliff2020adaptive, maheshwari2024adaptive} rely on two-time scale dynamics, where the game dynamics converge quickly, while incentives are updated at a slower timescale. However, in practice, game dynamics may converge slowly or not at all. To the best of our knowledge, few works have addressed the design of dynamic interventions for network games with continuous actions within a single-time scale dynamic framework. A case in point is \cite{shakarami2023dynamic}, which highlights that the suitable controller (namely, intervention protocol) depends on knowledge available to the regulator, and develops several intervention control policies tailored to the available knowledge of the regulator in strongly monotone network games. 

In this work, we address the problem of steering the actions of self-interested agents in quadratic network games toward a target action profile without access to the agents' private utility parameters or the interaction network. We model the decision-making process among agents using best-response dynamics, while the central regulator intervenes by modifying the agents' standalone marginal costs. Unlike \cite{shakarami2023dynamic, maheshwari2024adaptive}, we make no assumptions about the stability of the best-response dynamics.
We formulate the intervention design as a \textit{direct} data-driven control problem, where the regulator devises  the incentive design policy directly from historically observed agent actions and interventions, without performing any intermediate utility function or interaction network identification step. We analyze the interconnected regulator-agents dynamics on a single time scale and establish convergence of the actions to the target profile. Notably, since the intervention signal often corresponds to taxes or subsidies in practice, we also account for budget constraints to limit the amount of interventions during the steering process. In both cases, the intervention design procedures are formulated as linear matrix inequalities (LMIs), which can be efficiently solved. The technical derivations of these LMIs are inspired by recent works on direct data-driven control \cite{de2019formulas} along with \cite{seuret2024robust} and \cite{breschi2023data}.

The paper is organized as follows: Section \ref{problem_formulation} formulates the intervention problem in network games. Section \ref{Intervention_protocol} establishes the data-driven intervention protocol and provides its analytical convergence. In Section \ref{Intervention_with_constraints}, a modified intervention protocol is developed to account for budget constraints, and the effectiveness of the proposed protocols is demonstrated in Section \ref{Case_study} on Cournot competition of differentiated goods. The manuscript closes with conclusions in Section \ref{Conclusion}.

\textit{Notation and preliminaries}: Let $\mathbb{R}$ and $\mathbb{R}_{>0}$ be the sets of real numbers and positive real numbers, respectively. We use $\mathbbb{1}$($\mathbbb{0}$) to denote the vector/matrix with all elements equal to 1(0) and use $I$ as the identity matrix.  We include the dimension of these vectors/matrices as a subscript, whenever needed.  We use $\mathbb{R}^n$ and $\mathbb{R}^{n\times m}$ to denote the sets of $n$-dimensional real vectors and $n\times m$ real matrices, respectively. For a vector $v\in\mathbb{R}^n$ (or a matrix $M\in \mathbb{R}^{n \times m}$), $v_{(k)}$ (respectively, $M_{(k)}$) denotes its $k$th element (row). For a square matrix $M\in \mathbb{R}^{n \times n}$, $M^{-1}$, $M^\top$ and $\det(M)$ denote the inverse, transpose and determinant of the matrix $M$, respectively. For a matrix $M\succeq 0$, we denote the ellipsoid  $\{v\in\mathbb{R}^n\mid v^\top M v\leq 1\}$ by $\mathcal{E}(M,1)$. The notation $M\succ(\prec) 0$ means $M$ is positive (negative) definite.  For any matrices $A=A^\top$, $B$, $C=C^\top$ of appropriate dimensions, $\big[\begin{smallmatrix}
    A & * \\ B & C
    \end{smallmatrix}\big]$ denotes the symmetric matrix $\big[\begin{smallmatrix}
    A & B^\top \\ B & C
    \end{smallmatrix}\big]$, and $\diag(A,B,C)$ denotes the block diagonal matrix with $A$, $B$ and $C$ on the diagonal. Given a set $\mathcal{N}=\{1,2,...,N\}$, $\col(x_n)_{n\in\mathcal{N}}$ denotes the stacked vector obtained from $x_n$. The Kronecker product is denoted by $\otimes$.

\section{Problem formulation}\label{problem_formulation}
\subsection{Network games}
We consider a simultaneous-move game among a fixed number of agents $\mathcal{N}=\{1,...,N\}$. The agents interact repeatedly with a central regulator as well as with each other according to an underlying interaction network. We denote the adjacency matrix of this network by $P\in\mathbb{R}^{N\times N}$, where $P_{nm}\in [0,1]$ denotes the measure of the strength of the interaction between agents $n$ and $m$. We assume that the network has no self loop, thus $P_{nn}=0$ for all $n\in\mathcal{N}$. The goal of agent $n$ is to select an action $x_n\in\mathbb{R}$ to maximize a utility function $U_n(x_n,s_n(x),u_n)$, which depends on its own action $x_n$, the aggregate of its neighbors' actions
\begin{equation}
    s_n(x):=\sum_{m=1}^N P_{nm}x_m
\end{equation}
with $x=\col(x_n)_{n\in\mathcal{N}}$, and a scalar intervention $u_n$ which is determined by the central regulator. In this paper, we restrict our attention to linear quadratic utility functions of the form
\begin{equation}\label{payoff}
    U_n(x_n,s_n(x),u_n) = -\frac{1}{2} q_n x^2_n +  x_n\big(w_n s_n(x)+d_n\big) + x_n u_n 
\end{equation}
where the quadratic term $\frac{1}{2} q_n x^2_n$ with the constant $q_n\in\mathbb{R}_{>0}$ is the private cost of agent $n$. The marginal return from increasing the action $x_n$, i.e., $w_n s_n(x)+d_n$  depends both on the agent $n$'s standalone marginal return $d_n$ as well as the aggregate action $s_n(x)$. The parameter $w_n\in\mathbb{R}\backslash \{0\}$ captures the impact of neighbors aggregate actions $s_n(x)$ on the marginal return. Note that if for each agent $n$, $w_n>0$, this is a game of strategic complements, and if for each agent $n$, $w_n<0$, this is a game of strategic substitutes. The term $x_n u_n$ is included to capture the intervention of the central regulator in modifying the standalone marginal return $d_n$ to $d_n+u_n$. Network games with payoffs structured as described in \eqref{payoff} have been extensively analyzed in the literature to model peer influences in social and economic processes \cite{ballester2006s,parise2019variational}. This includes the particular case where $q_n=1$, as discussed in works such as \cite{galeotti2020targeting,parise2021analysis,shakarami2023dynamic}.

In network games, the agents are noncooperative and merely interested in maximizing their individual utility functions by choosing their actions. 
Given an intervention $u=\col(u_n)_{n\in\mathcal{N}}$,
an action profile $x^{ne}=\col(x_n^{ne})_{n\in\mathcal{N}}$ is a Nash equilibrium  if for all $x_n\in\mathbb{R}$ and $n\in\mathcal{N}$,
\begin{equation}
    U_n(x_n^{ne},s_n(x^{ne}),u_n)\geq  U_n(x_n,s_n(x^{ne}),u_n).
\end{equation} 
It is observed that at Nash equilibrium, no agent can improve its objective function by unilaterally changing its action. Taking the first-order derivative of the payoff $U_n$ with respect to the action $x_n$ in \eqref{payoff}, we have
\begin{equation}
    \partial_{x_n} U_n = -q_nx_n+w_n s_n(x)+d_n+u_n.
\end{equation}
Combining the above equations for all $n$, and letting $Q=\diag(q_n)_{n\in\mathcal{N}}$, $W=\diag(w_n)_{n\in\mathcal{N}}$, $d=\col(d_n)_{n\in\mathcal{N}}$, the following equality holds for any Nash equilibrium $x^{ne}$,
\begin{equation}\label{eq_NE}
    (Q-WP)x^{ne}-d-u=0,
\end{equation}
\subsection{Best response dynamics}
To maximize its individual utility, each agent iteratively updates its action based on the aggregated actions of its neighbours and the current value of the intervention signal. We assume that agents follow discrete-time best response dynamics. In particular, the action of each agent $n\in\mathcal{N}$ evolves according the following dynamics,
\begin{equation}\label{best_response1}
    x_n(k+1) = \argmax_{y_n} U_n\big(y_n,s_n(x(k)),u_n(k)\big).
\end{equation}
where $k$ indexes the time steps at which all agents simultaneously update their actions. We recall that $s_n(x(k))$ does not depend on $x_n(k)$ as $P_{nn}=0$, and $u_n(k)$ is the intervention designed by the regulator at time step $k$. The above best response dynamics are equivalent to
\begin{equation}\label{best_response2}
    x_n(k+1) = \frac{1}{q_n} w_n s_n(x(k))+\frac{1}{q_n}d_n +\frac{1}{q_n} u_n(k),
\end{equation}
which can be written in a compact form as 
\begin{equation}\label{compact_best_response}
    x(k+1) = Q^{-1}WP x(k) + Q^{-1}d +Q^{-1}u(k).
\end{equation}
The best response dynamics are commonly used in noncooperative games; see \cite{parise2019variational,scutari2014real}.
Note that for any constant $u=\bar u$, the best response dynamics \eqref{best_response2} admits a unique equilibrium if and only if the matrix $Q-WP$ is nonsingular. Such an equilibrium is given by 
\begin{equation}\label{eq_unique_ne}
\bar x= (Q-WP)^{-1}(d+\bar u)    
\end{equation}
and is a NE of the game as it satisfies \eqref{eq_NE}. 

\subsection{Target action profile}
Due to the fact that the self-interested behaviors of agents
may deviate or be in contrast with what is desired for the
group as a whole,  the central regulator is required to
coordinate the players by applying suitable interventions with
the aim of steering the players to a more desirable group
behavior \cite{galeotti2020targeting,parise2021analysis}. Here, we consider the scenario where the central regulator aims to incentivize the actions of agents to a desired setpoint $x^*$, which satisfies some agent-dependent constraints \cite{alpcan2009nash}, an aggregative constraint \cite{shakarami2022steering} or coincides with the solution of a social optimization problem \cite{maheshwari2024adaptive,ratliff2020adaptive}. 

By \eqref{eq_unique_ne}, the corresponding intervention at the equilibrium is given by 
\begin{equation}\label{eq_u_star}
  u^*=\left(Q-WP\right)x^*-d  
\end{equation}

We observe that implementing the above intervention requires the central regulator to know the utilities of the agents, the network structure, and the coupling weights. However, such information is typically not available a priori. To circumvent this challenge, we assume that the regulator can observe the actions of the players in a finite time interval and collect data. Moreover, the dynamics \eqref{compact_best_response} do not converge under a constant intervention unless the matrix $Q^{-1}WP$ is Schur stable. To cope with potential instability of this matrix, we use dynamic stabilizing  feedback protocols. Overall, the subsequent results address the following problem: 
\begin{problem}\label{pro}
    Design an intervention signal $u$ that steers the action profile $x$ of the agents to a target action profile $x^*$ without requiring the information of agents' utilities and the network parameters.
\end{problem}

\section{Intervention protocol}\label{Intervention_protocol}
We introduce the following PI controller as the intervention protocol executed by the central regulator:
\begin{subequations}\label{PI}
   \begin{align} \label{eq_z_dynamic}
    z(k+1)&= z(k) + x^* - x(k)\\
    u(k)&=K_xx(k)+ K_z z(k)
\end{align} 
\end{subequations}
where the matrices $K_x\in\mathbb{R}^{N\times N}$ and $K_z\in\mathbb{R}^{N\times N}$ are to be designed. 
Here, $z\in\mathbb{R}^N$ corresponds to the cumulative value of the differences between the target action profile $x^*$ and the actual action profile $x(k)$. 

Combining \eqref{PI} with the best response dynamics \eqref{compact_best_response}, we have the following closed-loop system,
%the discrete-time linear system,
\begin{subequations}\label{eq_closedloop}
\begin{align}\label{eq_dynamic}
    \begin{bmatrix}
        x(k+1) \\
        z(k+1)
    \end{bmatrix}=A  
    \begin{bmatrix}
        x(k) \\
        z(k)
    \end{bmatrix}
    +  B u(k) 
    +   E 
    \end{align}
    \vspace{-4mm}
    \begin{align}\label{eq_controller}
     u(k)
    = \underbrace{\begin{bmatrix}
        K_x &  K_z
    \end{bmatrix}}_{K} 
    \begin{bmatrix}
        x(k) \\
        z(k)
    \end{bmatrix},
    \end{align}
\end{subequations}
where
$$A =\begin{bmatrix}
        Q^{-1}WP & \mathbbb{0} \\
        -I & I
    \end{bmatrix}, B =\begin{bmatrix}
        Q^{-1} \\
        \mathbbb{0}
    \end{bmatrix}, E=\begin{bmatrix}
        Q^{-1}d \\
        x^*
    \end{bmatrix}.$$
Assuming that $K_z$ is nonsingular, the above closed-loop  
system admits a unique equilibrium $(x^*, z^*)$ with
\[
z^*= K_z^{-1}\Big( (Q-WP-K_x)x^*- d\Big).
\]
Therefore, at the equilibrium, the action profile coincides with the target profile as desired, and the problem reduces to designing $K_x$  and $K_z$ such that the closed-loop is asymptotically stable. 

As mentioned, we consider the scenario  where the regulator does not have a priori access to the utility functions of the agents and the information of the network, and instead has access to some historical data from the agent dynamics in \eqref{compact_best_response}. In particular, we assume that the regulator has collected a $T$-length intervention data sequence as $\{ u_d(t)\}_{t=0}^{t=T-1}$ and corresponding $(T+1)$-length action data sequence $\{x_d(t)\}_{t=0}^{t=T}$ obtained from \eqref{compact_best_response}. 

To state the main result of this section, let
\begin{equation}
   U_{[i,j]}: = \begin{bmatrix}
        u_d(i) & u_d(i+1) & \ldots  & u_d(j)
    \end{bmatrix},
\end{equation}
\begin{equation}
   X_{[i,j]}:= \begin{bmatrix}
        x_d(i) & x_d(i+1) & \ldots  & x_d(j)
    \end{bmatrix},
\end{equation}
and define the matrices
\begin{equation}\label{eq_data_matrix}
\begin{aligned}
V_0&=U_{[1,T-1]}-U_{[0,T-2]},\\
\Xi_0&=\begin{bmatrix}X_{[1,T-1]}-X_{[0,T-2]} \\ x^*\otimes \mathbbb{1}^{\top}_{T-1}- X_{[0,T-2]}\end{bmatrix},\\
\Xi_1&=\begin{bmatrix}X_{[2,T]}-X_{[1,T-1]} \\ x^*\otimes \mathbbb{1}^{\top}_{T-1}- X_{[1,T-1]} \end{bmatrix}.
\end{aligned}
\end{equation}

\begin{theorem}\label{theorem1}
Suppose $\Theta\in\mathbb{R}^{(T-1)\times 2N}$ satisfies the matrix inequality
\begin{equation}\label{eq_theta_matrix}
  \begin{bmatrix}
        \Xi_{0} \Theta & \Xi_{1} \Theta \\
        \Theta^{\top} \Xi_{1}^{\top}  & \Xi_{0} \Theta
\end{bmatrix} \succ 0.  
\end{equation}
Then, the intervention protocol \eqref{PI} with 
\begin{equation}\label{eq_theta_K}
 \begin{bmatrix}K_x &K_z \end{bmatrix}= V_{0} \Theta (\Xi_{0} \Theta)^{-1}   
\end{equation}
renders the equilibrium of \eqref{eq_closedloop} asymptotically stable. Consequently, the action profile asymptotically converges to the target profile $x^*$.
\end{theorem}

\begin{proof}
We define the auxiliary variables 
$$\xi(k):=\begin{bmatrix}
        x(k+1) \\
        z(k+1)
    \end{bmatrix}- \begin{bmatrix}
        x(k) \\
        z(k)
    \end{bmatrix},v(k):=u(k+1)-u(k).$$
Then, the dynamics in \eqref{eq_closedloop} transform into 
\begin{subequations}\label{eq_new_closedloop}
\begin{align}\label{eq_new_dynamic}
    \xi(k+1) = A  \xi (k) + B  v(k),
\end{align}
\begin{align}\label{eq_new_K}
    v(k) = K \xi(k).
\end{align}
\end{subequations}
The matrices $A$, $B$, and $K$ are the same as those in \eqref{eq_closedloop}. Analogously, the input-state data sequences $\{ u_d(t)\}_{t=0}^{t=T-1}$ and $\{ x_d(t)\}_{t=0}^{t=T}$ are mapped to
\begin{equation}\label{eq_v_data}
 \{v_d(t)\}_{t=0}^{t=T-2}, \quad v_d(t):=u_d(t+1)-u_d(t)   
\end{equation}
and
\begin{equation}\label{eq_xi_data}
   \{\xi_d(t)\}_{t=0}^{t=T-1}, \quad \xi_d(t):=\begin{bmatrix}
    x_d(t+1)-x_d(t) \\ x^*-x_d(t)
\end{bmatrix},
\end{equation}
respectively. Note that we used \eqref{eq_z_dynamic} to write the expression of $\xi_d(t)$.
Then, we observe that the matrices defined in \eqref{eq_data_matrix}, in fact, correspond to the input-state data of the auxiliary system  \eqref{eq_new_closedloop}. In particular, 
\begin{equation}\label{eq_data}
    \Xi_1=A\Xi_0+B V_0,
\end{equation}
with
\[
V_0=\begin{bmatrix}
    v_d(0) & v_d(1) &... & v_d(T-2)
\end{bmatrix},
\]
\[
\Xi_0=\begin{bmatrix}
    \xi_d(0) & \xi_d(1) &... & \xi_d(T-2)
\end{bmatrix}, 
\]
\[
\Xi_1=\begin{bmatrix}
    \xi_d(1) & \xi_d(2) &... & \xi_d(T-1)
\end{bmatrix}.
\]
By the proof of \cite[Thm. 3]{de2019formulas}, it follows that the controller obtained from \eqref{eq_theta_matrix}-\eqref{eq_theta_K} stabilizes the system in \eqref{eq_new_closedloop}, implying that the matrix $A+BK$ is Schur stable. This, together with the fact that the matrix $E$ is constant proves asymptotic stability of the equilibrium in \eqref{eq_closedloop}, and asymptotic convergence of the action profiles to $x^*$ \cite{hespanha2018linear}.
\end{proof}

\begin{remark}\label{remark1}
If the data sequence $\{v_d(t)\}_{t=0}^{t=T-2}$ in \eqref{eq_v_data} is Persistently Exciting (PE) of order $2N+1$, then a matrix $\Theta$ satisfying \eqref{eq_theta_matrix} always exists. To see this, first note that the the pair $(A, B)$ is controllable as the following matrix, 
\begin{equation}
    \begin{bmatrix}
        B & AB
    \end{bmatrix}=
    \begin{bmatrix}
        Q^{-1} & Q^{-1} WPQ^{-1}  \\
        \mathbbb{0} & -Q^{-1} 
    \end{bmatrix}.
\end{equation}
has full row rank. Then, the model-based stabilization problem of the auxiliary system \eqref{eq_new_closedloop} has a solution. Moreover, the matrix $\big [\begin{smallmatrix}
    V_0 \\ \Xi_0
\end{smallmatrix}\big]$ has full row rank \cite{willems2005note}, which ensures the data-based stabilization problem of \eqref{eq_new_closedloop} has a solution in the form of \eqref{eq_theta_K}; see the discussion following \cite[Thm. 3]{de2019formulas}. 
\end{remark}

\begin{remark}\label{remark11}
The computational complexity of solving \eqref{eq_theta_matrix} is roughly  $\mathcal{O}(N^6)$,  which is comparable to a sequential system identification and controller design; see the discussion on direct vs. indirect data-driven control \cite[Sec. 2]{sznaier2020control}.
The key advantage of a direct data-driven control method, as in the case of Theorem \ref{theorem1}, is that, in \textit{one shot}, the controller and the Lyapunov function certifying stability are obtained from data. As a numerical consideration, since $\Xi_0 {\Theta}$ cannot be directly interpreted as a symmetric matrix by numerical solvers such as CVX, we can replace $\Xi_{0} \Theta$ in the diagonal blocks of \eqref{eq_theta_matrix} by a new variable $\Theta_{in}$, and add an equality constraint $\Theta_{in}=\Xi_{0} \Theta$ to account for this variable change. 
\end{remark}

\section{Intervention with budget constraints}\label{Intervention_with_constraints}
As the intervention signal amounts to tax or subsidies in practice, it is natural to impose a budget constraint on the level of intervention. To this end, we consider the set
\begin{equation}\label{set_u}
  \mathcal{C}(u):=\{u\in\mathbb{R}^N \mid u_{\min} \leq u_n \leq u_{\max}, \, n\in\mathcal{N}\} 
\end{equation}
where $u_{\min}\in\mathbb{R}$ and $u_{\max}\in\mathbb{R}$ are the minimum and maximum allowed intervention performed by the regulator for each agent.

In this section, we are interested in solving Problem \ref{pro} with $u(k)\in \mathcal{C}(u)$ for all $k\geq 0$.
As a necessary condition for solvability of this problem, we assume that at the desired equilibrium, namely $x^*$, the budget constraint is strictly satisfied for each agent: 
\begin{assumption}\label{assumption1}
Given a target action profile $x^*$, the corresponding $u^*$ in \eqref{eq_u_star} strictly satisfies $u^*\in \mathcal{C}(u)$.
%the intervention constraint, namely, $u^*\in \mathcal{C}(u)$.
\end{assumption}

Before proceeding further, we define the saturation operator as 
\[
    \sat_{a}^b(\tau)= \begin{cases}
    b & \tau > b \\
   \tau & a \leq \tau \leq b\\
    a & \tau < a,
    \end{cases}
\]
for given scalars $a$ and $b$.
For a vector $\tau$, the operator is applied element-wise.

Taking into account budget constraint, we modify the PI control \eqref{PI} into the following intervention protocol,
\begin{subequations}\label{PI1}
   \begin{align} \label{eq_z_dynamic1}
    z(k+1)&= z(k) + x^* - x(k), \\
    u(k)& = \sat_{u_{\min}}^{u_{\max}}\big(K_xx(k)+ K_z z(k)\big).
\end{align} 
\end{subequations}
This, together with the best response dynamics \eqref{compact_best_response}, results in the closed-loop system 
\begin{equation}\label{eq_closedloop_cons}
    \begin{bmatrix}
        x(k+1) \\
        z(k+1)
    \end{bmatrix}=A  
    \begin{bmatrix}
        x(k) \\
        z(k)
    \end{bmatrix}
    +  B \sat_{u_{\min}}^{u_{\max}}\big(K_xx(k)+ K_z z(k)\big)
    +   E. 
\end{equation}
We shift the equilibrium to the origin by defining a new state variable $\delta:=\begin{bmatrix}
        (x-x^*)^\top &
        (z-z^*)^\top
    \end{bmatrix}^\top$, which results in
\begin{equation}\label{eq_constrain_new1}
    \delta(k+1)=A \delta(k)+ B \sat_{u_{\min}-u^*}^{u_{\max}-u^*}(K\delta(k)).
\end{equation}
The closed loop system \eqref{eq_constrain_new1} can further be written as
\begin{equation}\label{eq_constrain_new2}
    \delta(k+1)=(A+BK)\delta(k)+B\phi(K\delta(k))
\end{equation}
with the dead-zone function defined as
\begin{equation}
    \phi(K\delta) =  \sat_{u_{\min}-u^*}^{u_{\max}-u^*}(K\delta)-K\delta.
\end{equation}
The above expression allows us to apply the following generalized sector condition; see \cite[Lem. 1.6]{tarbouriech2011stability}) for a proof.
\begin{lemma}\label{lemma1} 
For any matrix $H\in\mathbb{R}^{N\times 2N}$ and any diagonal positive definite matrix $S\in\mathbb{R}^{N\times N}$, we have
    \begin{equation}\label{eq_section_condition}
        \phi^\top(K\delta)S^{-1}(\phi(K\delta)+K\delta+H\delta)\leq 0
    \end{equation}
for all $\delta \in \mathcal{S}_H$, with 
    \begin{equation}\label{set_s}
        \mathcal{S}_H(\delta):=\{\delta\in\mathbb{R}^{2N} \mid \, |H_{(n)}\delta| \leq \rho_n,  n\in\mathcal{N}\}
    \end{equation}
and
\begin{equation}\label{ubar-min}
\rho_n := \min\{|u_n^*-u_{\min}|,|u_{\max}-u_n^*|\}.
\end{equation}
\end{lemma}

The inclusion of the budget constraints complicates the data-driven design. To partially tame this complexity, in the sequel, we assume the case that $Q=I_N$ as done in \cite{galeotti2020targeting,parise2021analysis,shakarami2023dynamic}. Consequently, the input matrix reduces to  $B=\begin{bmatrix} I_{N} & \mathbbb{0}_{N\times N} \end{bmatrix}^\top$. Bearing in mind the data matrices $V_0$, $\Xi_0$ and $\Xi_1$ in \eqref{eq_data_matrix}, we have the following result.

\begin{theorem}\label{theorem2}
Under Assumption \ref{assumption1}, suppose $\bar\Theta\in\mathbb{R}^{(T-1)\times 2N}$, a diagonal positive definite matrix $S\in \mathbb{R}^{N\times N}$, and $Z\in\mathbb{R}^{N\times 2N}$ satisfy the matrix inequalities
\begin{equation}\label{eq_theorem2_matrix1}
    \begin{bmatrix}
         \Xi_{0} \bar\Theta &   *  & * \\ 
        V_0\bar\Theta+Z & 2S & * \\
        \Xi_1 \bar\Theta  & \begin{bmatrix}
    I_{N} & \mathbbb{0}_{N\times N}
\end{bmatrix}^\top S & \ \Xi_{0} \bar\Theta
        \end{bmatrix}\succ 0
\end{equation}
and 
\begin{equation}\label{eq_theorem2_matrix2}
\begin{bmatrix}
    \Xi_{0} \bar\Theta & Z_{(n)}^\top \\
    Z_{(n)}  & \rho_n^2
\end{bmatrix}\succ 0,
\end{equation}
 for each $n\in\mathcal{N}$.
Let the intervention protocol \eqref{PI1} be designed with gains 
\begin{equation}\label{eq_theta_K2}
 \begin{bmatrix}K_x &K_z \end{bmatrix}= V_0 \bar\Theta (\Xi_{0} \bar\Theta)^{-1}.
\end{equation}
Then, the equilibrium $(x^*, z^*)$ of the closed-loop system \eqref{eq_closedloop_cons} is asymptotically stable. Moreover, an estimate of the region of attraction is given by the set
\begin{equation}\label{ellipsoid}
 \mathcal{R}:=
\{(x, z)\mid 
 \big[\begin{smallmatrix}
        x-x^* \\ 
        z-z^*
    \end{smallmatrix}\big] \in \mathcal{E}((\Xi_{0} \bar\Theta)^{-1},1)
 \}.   
\end{equation}
\end{theorem}

\begin{proof}
Consider the Lyapunov candidate for the system \eqref{eq_constrain_new2} as 
$V(\delta)=\delta^\top M \delta$ with a positive definite matrix $M$. The forward increment of the Lyapunov candidate is given by  
$V(\delta(k+1))- V(\delta(k))=\gamma^\top(k)\Gamma_1\gamma(k) $
where
\begin{equation}
\Gamma_1:=\begin{bmatrix}
       (A+BK)^\top \\
       B^\top \end{bmatrix}M
      \begin{bmatrix}
       (A+BK)^\top \\
       B^\top \end{bmatrix}^\top-\begin{bmatrix}
       M & 0\\
      0 & 0 \end{bmatrix},
\end{equation}
and $\gamma(k)=[\delta^\top(k)  \;\; \phi^\top(K\delta(k))]^\top$.
We drop the index $k$ in the rest of the proof for notational simplicity.
To ensure asymptotic stability of the equilibrium, it suffices to have
\begin{equation}\label{eq_stability}
\gamma^\top\Gamma_1\gamma\leq 0, \quad \forall \, \delta \in \mathcal{E}(M,1). 
\end{equation}
By leveraging the generalized sector condition in Lemma \ref{lemma1}, a sufficient condition for \eqref{eq_stability} is provided in the proof of \cite[Thm. 1]{seuret2024robust}. Namely, \eqref{eq_stability} holds if 
\begin{equation}\label{gamma2}
    \Gamma_{2,n}:=\begin{bmatrix}
    M^{-1} & * \\
    H_{(n)} M^{-1} & \rho_n^2
\end{bmatrix}\succ 0
\end{equation}
for each $n\in\mathcal{N}$, and
\begin{equation}\label{Gamma3}
\begin{aligned}
   \Gamma_3: =& \begin{bmatrix}
        M &   * \\ 
        S^{-1}(K+H) & 2S^{-1}
        \end{bmatrix}\\
        -& \begin{bmatrix}
       (A+BK)^\top \\
       B^\top \end{bmatrix}M
      \begin{bmatrix}
       (A+BK)^\top \\
       B^\top \end{bmatrix}^\top \succ 0,
\end{aligned}
\end{equation}
for some matrices $H\in \mathbb{R}^{N\times 2N}$, $S\succ 0$ and diagonal, and $\rho_n$ given by \eqref{ubar-min}. By viewing $\Gamma_3$ as a Schur complement of a ($3\times 3$ block) matrix, and pre- and post-multiplying that matrix by $\diag(M^{-1},S,I_{2N})$, we find that 
\eqref{Gamma3} is equivalent to
\begin{equation}\label{matrix_afer_multiply}
     \Gamma_4=\begin{bmatrix}
        M^{-1} &  * & * \\ 
        (K+H)M^{-1} & 2S & * \\
        (A+BK)M^{-1} & BS & M^{-1}
        \end{bmatrix}\succ 0.
\end{equation}
Next, recall that the data matrices $V_0$, $\Xi_0$ and $\Xi_1$  satisfy \eqref{eq_data}, and that $B=\begin{bmatrix} I_{N} & \mathbbb{0}_{N\times N} \end{bmatrix}^\top$. Bearing in mind \eqref{eq_theta_K2}, multiplying both sides of \eqref{eq_data} by $\bar\Theta (\Xi_0 \bar\Theta)^{-1}$ yields $A+BK=\Xi_1  \bar\Theta (\Xi_{0} \bar\Theta)^{-1}$. Now, choosing $M:=(\Xi_0 \bar\Theta)^{-1}$ and substituting the values of $A+BK$ from the latter equality and $K$ from \eqref{eq_theta_K2} into \eqref{matrix_afer_multiply} result in \eqref{eq_theorem2_matrix1}, where a change of variable $Z=H\Xi_0\bar\Theta$ is also carried out to obtain an LMI. In addition, \eqref{gamma2} becomes equal to \eqref{eq_theorem2_matrix2}. 
This completes the proof for asymptotic stability of the equilibrium.
The claim concerning the region of attraction follows from \eqref{eq_stability} and the definition of $\delta$ variables in \eqref{eq_constrain_new1}.
\end{proof}

The next result addresses the feasibility of the LMIs established in Theorem \ref{theorem2}.

\begin{lemma}\label{lemma2}
If the data sequence $\{v_d(t)\}_{t=0}^{t=T-2}$ in \eqref{eq_v_data} is persistently exciting of order $2N+1$, then 
the LMIs
 \eqref{eq_theorem2_matrix1}- \eqref{eq_theorem2_matrix2} have a feasible solution. 
\end{lemma}

\begin{proof}
Since the pair $(A, B)$ in \eqref{eq_new_closedloop} is controllable, the model-based stabilization problem of the auxiliary system \eqref{eq_new_closedloop} has a solution. That is, there always exist a positive definite matrix $\bar M\succ 0$ and a controller $\bar K$ such that 
\begin{equation}\label{R1}
        \begin{bmatrix}
         \bar M^{-1} &  *\\
       (A+B\bar K)\bar M^{-1} & \bar M^{-1}
    \end{bmatrix} \succ 0.
\end{equation}
Note that taking the Schur complement of the matrix above yields the Lyapunov inequality $ \bar M^{-1}-(A+B\bar K)\bar M^{-1}(A+ B \bar K)^\top \succ 0$, certifying closed-loop stability. 

Next, we show that the model-based matrix inequalities \eqref{gamma2} and \eqref{matrix_afer_multiply} have a solution. To this end, we choose $M^{-1}=\alpha \bar M^{-1}$, $K=\bar K$, and $H=-\bar K$, with some positive scalar $\alpha$ to be specified later. 
 Then the matrix $\Gamma_{2,n}$ in \eqref{gamma2} becomes
\begin{equation*}
    \Gamma_{2,n}=\begin{bmatrix}
    \alpha \bar M^{-1} & * \\
    -\alpha \bar K_{(n)} \bar M^{-1} & \rho_n^2
\end{bmatrix}
\end{equation*}
and the matrix $\Gamma_4$ in \eqref{matrix_afer_multiply}, after applying block row and column permutations, simplifies to 
\begin{equation*}
     \Gamma_4=\begin{bmatrix}
        \alpha \bar M^{-1} &  * & * \\ 
        \alpha(A+B\bar K)\bar M^{-1} & \alpha \bar M^{-1} & * \\
        0 &  S B^\top & 2S
        \end{bmatrix}.
\end{equation*}
By using a Schur complement argument, we find that $\Gamma_{2,n} \succ 0$ if and only if $\rho_n^2> \alpha K_{(n)} \bar M^{-1} K^\top_{(n)}$. We select a sufficiently small $\alpha$ such that the latter inequality holds and thus $\Gamma_{2,n}\succ 0$. 

Moreover, by using again a Schur complement argument, we have $\Gamma_4\succ 0$ if and if 
\begin{equation}\label{R2}
\alpha \begin{bmatrix}
         \bar M^{-1} &  *\\
       (A+B\bar K)\bar M^{-1} & \bar M^{-1}
    \end{bmatrix}- \frac{1}{2}\begin{bmatrix}
        0 &  0\\
       0 & BSB^\top
    \end{bmatrix} \succ 0.
\end{equation}
By \eqref{R1}, the first matrix on the left is positive definite. Therefore, there 
always exist a positive definite diagonal matrix $\bar S$ satisfying the above inequality. We conclude that the model-based matrix inequalities \eqref{gamma2} and \eqref{matrix_afer_multiply} have a solution  $(M,K,H,S)=(\alpha^{-1} \bar M, \bar K, - \bar K, \bar S)$, with $S=\bar S$ satisfying \eqref{R2}. 

To complete the proof, it suffices to show that if the matrix inequalities \eqref{gamma2} and \eqref{matrix_afer_multiply} have a solution, then also the data-based LMIs \eqref{eq_theorem2_matrix1}-\eqref{eq_theorem2_matrix2} admit a solution. To this end, let $(M_m, K_m, H_m, S_m)$ be any feasible solution to \eqref{gamma2} and \eqref{matrix_afer_multiply}. By using Theorem 2 in \cite{de2019formulas}, if the data sequence $\{v_d(t)\}_{t=0}^{t=T-2}$ in \eqref{eq_v_data} is persistently exciting of order $2N+1$, there exists $G_K$ such that 
\begin{equation}\label{R3}
    A+BK_m=\Xi_1 G_K,
\end{equation}
\begin{equation}\label{R4}
    K_m = V_0 G_K, \quad 
    I = \Xi_0 G_K.
\end{equation}
By choosing $\bar \Theta =G_K M_m^{-1}$, $Z=H_m\Xi_0 \bar\Theta$ and $S=S_m$, the data-based LMIs \eqref{eq_theorem2_matrix1}-\eqref{eq_theorem2_matrix2}  
become

\begin{equation*}
\footnotesize{    \begin{bmatrix}
         \Xi_{0}G_K M_m^{-1}  &   *  & * \\ 
        V_0G_K M_m^{-1}+H_m\Xi_0G_K M_m^{-1} & 2S_m & * \\
        \Xi_1 G_K M_m^{-1}  & \begin{bmatrix}
    I_{N} & \mathbbb{0}_{N\times N}
\end{bmatrix}^\top S_m & \ \Xi_{0}G_K M_m^{-1}
        \end{bmatrix} \succ 0,}
\end{equation*}

\begin{equation*}
\begin{bmatrix}
    \Xi_{0}G_K M_m^{-1}  & * \\
    (H_m\Xi_0 \bar\Theta)_{(n)}  & \rho_n^2
\end{bmatrix}\succ 0.
\end{equation*}
By using the equalities \eqref{R3} and \eqref{R4}, 
we observe that the above inequalities coincide with the model-based matrix inequalities \eqref{gamma2} and \eqref{matrix_afer_multiply} and thus are satisfied. This completes the proof. 
\end{proof}

\begin{remark}\label{remark2}
The LMIs in \eqref{eq_theorem2_matrix1}-\eqref{eq_theorem2_matrix2} can be solved directly using the data matrices in \eqref{eq_data_matrix}, returning the controller, the Lyapunov function certifying stability, and an estimate of the region of attraction. Similar to Remark \ref{remark1}, the LMIs \eqref{eq_theorem2_matrix1}-\eqref{eq_theorem2_matrix2} are feasible under the PE condition, as proven in Lemma \ref{lemma2}. We also note that the only parameter that is not readily available from data is $\rho_n$, given by \eqref{ubar-min}. To remedy this, the regulator can leverage knowledge on the bounds of the model parameters  to obtain bounds on $u^*$ in \eqref{eq_u_star}. Such bounds can in turn be used to find a lower bound of $\rho_n$. Although the feasibility of the proposed LMIs remains unaffected if $\rho_n$ is replaced by any lower bound $\hat{\rho}_n \leq \rho_n$, the quality of this lower bound could potentially shrink the size of the estimated region of attraction, as this estimate is a subset of $\mathcal{S}_H$ in \eqref{set_s}.
\end{remark}

\begin{remark}
The volume of the ellipsoid in \eqref{ellipsoid} is proportional to $\det(\Xi_0 \bar\Theta)$. Given that this value as a result of solving \eqref{eq_theorem2_matrix1} and \eqref{eq_theorem2_matrix2} may be relatively small, one can add an additional inequality $\Xi_{0} \bar\Theta \succ \eta I_{2N}$ and maximize $\eta$ subject to the LMIs \eqref{eq_theorem2_matrix1} and \eqref{eq_theorem2_matrix2}. 
\end{remark}

\section{Case study}\label{Case_study}
\begin{figure}[!ht]
\centering
\includegraphics[width=0.9\linewidth]{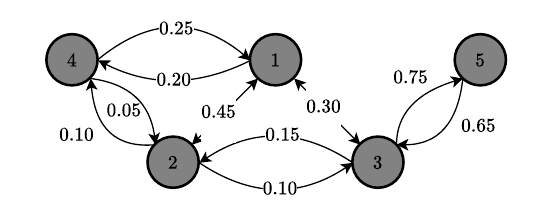}
\vspace*{-0.5cm}
\caption{The direct graph illustrating asymmetrical product substitutability}
\label{graph}
\end{figure}

We consider a Cournot competition involving a set of firms, $\mathcal{N} = \{1, 2, \dots, 5\}$, that produce differentiated goods \cite{bramoulle2014strategic}. For each firm $n$, we denote the amount of goods by $x_n$, and its corresponding price is obtained from the inverse demand function $p_n(x)=\alpha_n-(\ell_n x_n+\beta_n\sum_{m\neq n}P_{nm}x_m)$, where $\alpha_n$ is the maximum price that consumers would pay for the good, $\beta_n$ and $\ell_n$ are the price elasticity coefficients, and $P_{nm}$ indicates the degree of product substitutability. 
The products of the firms are substituable according to the weighted directed graph depicted in Fig. \ref{graph}, where the weight on the link from node $m$ to node $n$ is equal to $P_{nm}$. The cost of good production is modeled as a linear quadratic function, i.e., $\frac{1}{2}a_n x^2_n+b_n x_n$ and the intervention is $u_nx_n$. Then, the payoff function of firm $n$ is written in the form of as
\eqref{payoff}
\begin{equation*}
  \begin{aligned}
     U_n(x_n,s_n(x),u_n) & = -\frac{1}{2} (a_n+\ell_n) x^2_n \\
     & + x_n\big(-\beta_n s_n(x)+\alpha_n-b_n\big) + x_n u_n .
\end{aligned}  
\end{equation*}
Thus, $q_n=a_n+2\ell_n$, $w_n=-\beta_n$, $d_n=\alpha_n-b_n$ and $s_n(x)=\sum_{m=1}^N P_{nm}x_m$.  
The numerical values of the parameters are given by 
$Q=\diag(q_n)=\diag(1, 1.2, 0.8, 1.4, 0.9)$, $d=\col(d_n)=\col(12, 8, 10, 6, 10)$, and $w_n=-1.8$ for $n\in\mathcal{N}$. We consider that the regulator has access to some historical 
intervention-action data $\{ u_d(t)\}_{t=0}^{t=T-1}, \{ x_d(t)\}_{t=0}^{t=T}$. For numerical implementation, we pick the intervention points $u_d(t)$ arbitrarily from a bounded interval $[-2, 2]$, and set $T=80$. In practice, any sequence of historical interventions satisfying the PE condition in Remark \ref{remark1} can be used. The central regulator sets the target production profile $x^*=[4.0, 5.2, 5.0, 7.0, 7.5]$, obtained from some fairness or  sustainability considerations. 

\begin{figure}[htbp!]
\centering
\includegraphics[width=0.95\linewidth]{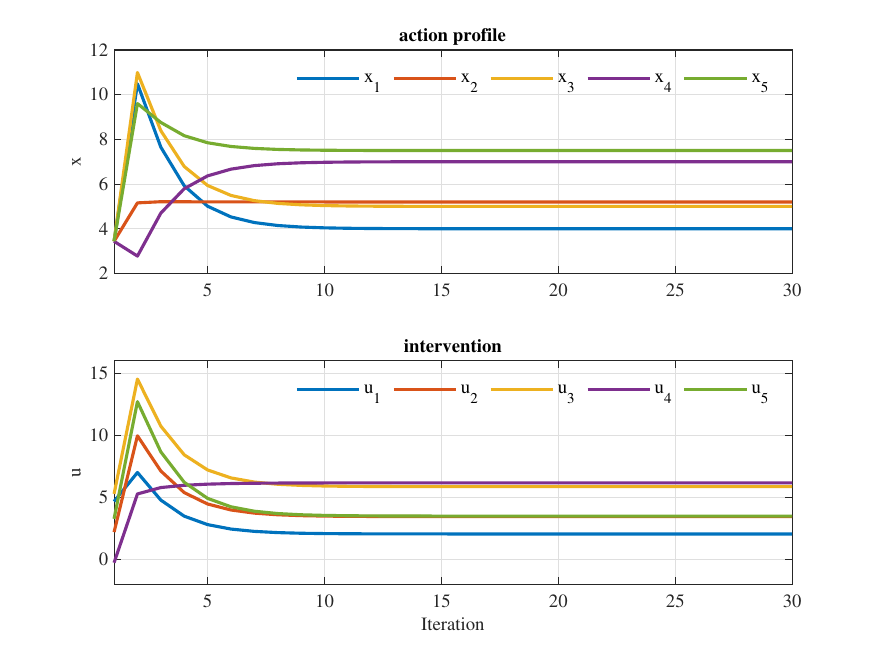}
%\vspace*{-5mm}
\caption{Agent action profile and intervention without budget constraints}
\label{protocol1}
\end{figure}

\begin{figure}[htbp!]
\centering
\includegraphics[width=0.95\linewidth]{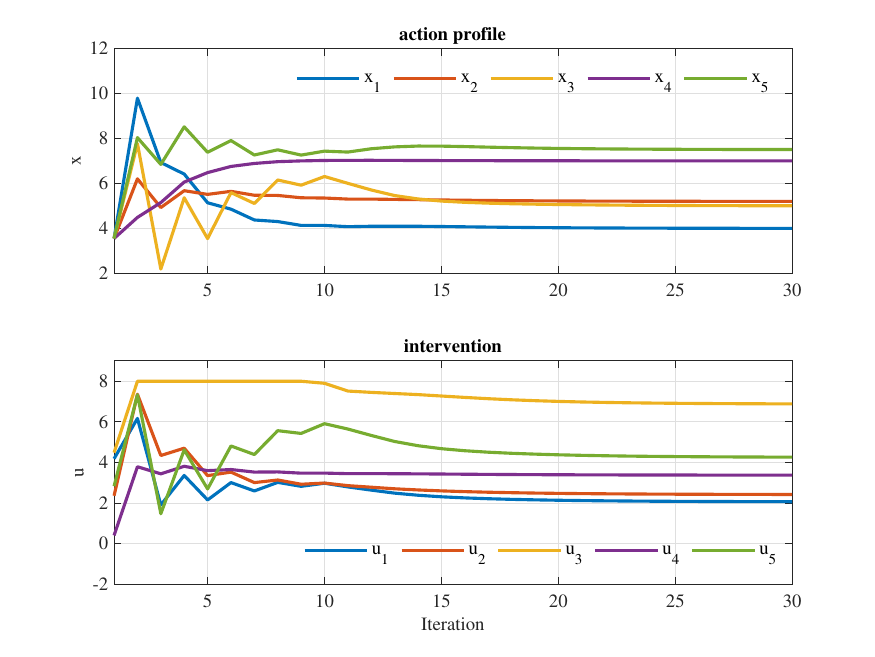}
%\vspace*{-5mm}
\caption{Agent action profile and intervention with budget constraints}
\label{protocol2}
\end{figure}

First, we consider the scenario without budget constraints and implement the proposed dynamic intervention in \eqref{PI}, with the control gain obtained from Theorem \ref{theorem1}.
Fig. \ref{protocol1} demonstrates the evolution of the intervention (lower plot) and the corresponding action profile (upper plot). As expected, the actions asymptotically converge to the target profile $x^*$. Next, we include the budget  constraint $\mathcal{C}(u)$ in \eqref{set_u} with $u_{\max}=-u_{\min}=8$, and set $Q=I_5$, and follow the data-driven design protocol in Theorem \ref{theorem2}. In  Figure \ref{protocol2}, the action profiles are again steered to the target profile. Moreover, the amount of intervention remains within the specified budget.

\section{Conclusion}\label{Conclusion}
We have adopted a data-driven approach to the design of intervention protocols in network games. This approach allows us to bypass the common assumption that utility functions and/or network parameters are available to the regulator a priori. In particular, we have simplified the problem of intervention design into a set of LMIs that depend on action-intervention data. The designed intervention protocols are capable of steering the outcomes of strategic interactions in network games toward a target action profile. Limitations on available resources for interventions are accounted for by enforcing a budget constraint on the devised dynamic protocols. The analytical results are complemented by a numerical case study demonstrating the effectiveness of the proposed design. Future work includes extending the results to network games with nonlinear dynamics, and more general game settings.

\bibliographystyle{IEEEtran} 
\bibliography{ref}

\begin{thebibliography}{10}
\providecommand{\url}[1]{#1}
\csname url@samestyle\endcsname
\providecommand{\newblock}{\relax}
\providecommand{\bibinfo}[2]{#2}
\providecommand{\BIBentrySTDinterwordspacing}{\spaceskip=0pt\relax}
\providecommand{\BIBentryALTinterwordstretchfactor}{4}
\providecommand{\BIBentryALTinterwordspacing}{\spaceskip=\fontdimen2\font plus
\BIBentryALTinterwordstretchfactor\fontdimen3\font minus \fontdimen4\font\relax}
\providecommand{\BIBforeignlanguage}[2]{{%
\expandafter\ifx\csname l@#1\endcsname\relax
\typeout{** WARNING: IEEEtran.bst: No hyphenation pattern has been}%
\typeout{** loaded for the language `#1'. Using the pattern for}%
\typeout{** the default language instead.}%
\else
\language=\csname l@#1\endcsname
\fi
#2}}
\providecommand{\BIBdecl}{\relax}
\BIBdecl

\bibitem{ballester2006s}
C.~Ballester, A.~Calv{\'o}-Armengol, and Y.~Zenou, ``Who's who in networks. wanted: The key player,'' \emph{Econometrica}, vol.~74, no.~5, pp. 1403--1417, 2006.

\bibitem{bloch2013pricing}
F.~Bloch and N.~Qu{\'e}rou, ``Pricing in social networks,'' \emph{Games and economic behavior}, vol.~80, pp. 243--261, 2013.

\bibitem{bramoulle2007public}
Y.~Bramoull{\'e} and R.~Kranton, ``Public goods in networks,'' \emph{Journal of Economic theory}, vol. 135, no.~1, pp. 478--494, 2007.

\bibitem{parise2019variational}
F.~Parise and A.~Ozdaglar, ``A variational inequality framework for network games: Existence, uniqueness, convergence and sensitivity analysis,'' \emph{Games and Economic Behavior}, vol. 114, pp. 47--82, 2019.

\bibitem{koutsoupias2009worst}
E.~Koutsoupias and C.~Papadimitriou, ``Worst-case equilibria,'' \emph{Computer science review}, vol.~3, no.~2, pp. 65--69, 2009.

\bibitem{galeotti2020targeting}
A.~Galeotti, B.~Golub, and S.~Goyal, ``Targeting interventions in networks,'' \emph{Econometrica}, vol.~88, no.~6, pp. 2445--2471, 2020.

\bibitem{parise2021analysis}
F.~Parise and A.~Ozdaglar, ``Analysis and interventions in large network games,'' \emph{Annual Review of Control, Robotics, and Autonomous Systems}, vol.~4, no.~1, pp. 455--486, 2021.

\bibitem{alpcan2009control}
T.~Alpcan, L.~Pavel, and N.~Stefanovic, ``A control theoretic approach to noncooperative game design,'' in \emph{Proceedings of the 48h IEEE Conference on Decision and Control (CDC) held jointly with 2009 28th Chinese Control Conference}.\hskip 1em plus 0.5em minus 0.4em\relax IEEE, 2009, pp. 8575--8580.

\bibitem{eksin2020control}
C.~Eksin and K.~Paarporn, ``Control of learning in anticoordination network games,'' \emph{IEEE Transactions on Control of Network Systems}, vol.~7, no.~4, pp. 1823--1835, 2020.

\bibitem{riehl2018incentive}
J.~Riehl, P.~Ramazi, and M.~Cao, ``Incentive-based control of asynchronous best-response dynamics on binary decision networks,'' \emph{IEEE Transactions on Control of Network Systems}, vol.~6, no.~2, pp. 727--736, 2018.

\bibitem{ratliff2020adaptive}
L.~J. Ratliff and T.~Fiez, ``Adaptive incentive design,'' \emph{IEEE Transactions on Automatic Control}, vol.~66, no.~8, pp. 3871--3878, 2020.

\bibitem{maheshwari2024adaptive}
C.~Maheshwari, K.~Kulkarni, M.~Wu, and S.~Sastry, ``Adaptive incentive design with learning agents,'' \emph{arXiv:2405.16716}, 2024.

\bibitem{shakarami2023dynamic}
M.~Shakarami, A.~Cherukuri, and N.~Monshizadeh, ``Dynamic interventions with limited knowledge in network games,'' \emph{IEEE Transactions on Control of Network Systems}, 2023.

\bibitem{de2019formulas}
C.~De~Persis and P.~Tesi, ``Formulas for data-driven control: Stabilization, optimality, and robustness,'' \emph{IEEE Transactions on Automatic Control}, vol.~65, no.~3, pp. 909--924, 2019.

\bibitem{seuret2024robust}
A.~Seuret and S.~Tarbouriech, ``Robust data-driven control design for linear systems subject to input saturation,'' \emph{IEEE Transactions on Automatic Control}, 2024.

\bibitem{breschi2023data}
V.~Breschi, L.~Zaccarian, and S.~Formentin, ``Data-driven stabilization of input-saturated systems,'' \emph{IEEE Control Systems Letters}, vol.~7, pp. 1640--1645, 2023.

\bibitem{scutari2014real}
G.~Scutari, F.~Facchinei, J.-S. Pang, and D.~P. Palomar, ``Real and complex monotone communication games,'' \emph{IEEE Transactions on Information Theory}, vol.~60, no.~7, pp. 4197--4231, 2014.

\bibitem{alpcan2009nash}
T.~Alpcan and L.~Pavel, ``Nash equilibrium design and optimization,'' in \emph{2009 international conference on game theory for networks}.\hskip 1em plus 0.5em minus 0.4em\relax IEEE, 2009, pp. 164--170.

\bibitem{shakarami2022steering}
M.~Shakarami, A.~Cherukuri, and N.~Monshizadeh, ``Steering the aggregative behavior of noncooperative agents: a nudge framework,'' \emph{Automatica}, vol. 136, p. 110003, 2022.

\bibitem{hespanha2018linear}
J.~P. Hespanha, \emph{Linear systems theory}.\hskip 1em plus 0.5em minus 0.4em\relax Princeton university press, 2018.

\bibitem{willems2005note}
J.~C. Willems, P.~Rapisarda, I.~Markovsky, and B.~L. De~Moor, ``A note on persistency of excitation,'' \emph{Systems \& Control Letters}, vol.~54, no.~4, pp. 325--329, 2005.

\bibitem{sznaier2020control}
M.~Sznaier, ``Control oriented learning in the era of big data,'' \emph{IEEE Control Systems Letters}, vol.~5, no.~6, pp. 1855--1867, 2020.

\bibitem{tarbouriech2011stability}
S.~Tarbouriech, G.~Garcia, J.~M.~G. da~Silva~Jr, and I.~Queinnec, \emph{Stability and stabilization of linear systems with saturating actuators}.\hskip 1em plus 0.5em minus 0.4em\relax Springer Science \& Business Media, 2011.

\bibitem{bramoulle2014strategic}
Y.~Bramoull{\'e}, R.~Kranton, and M.~D'amours, ``Strategic interaction and networks,'' \emph{American Economic Review}, vol. 104, no.~3, pp. 898--930, 2014.

\end{thebibliography}
\end{document}